\documentclass[prb,aps,twocolumn, showpacs]{revtex4}

\usepackage{graphicx}
\usepackage{color}

\def\v#1{\mathbf{#1}}

\begin{document}

\title{Fundamental collapse of the exciton-exciton effective scattering}
\author{Laura Pilozzi$^{1}$, Monique Combescot$^{2}$, Odile 
Betbeder-Matibet$^{2}$, Andrea D'Andrea$^{1}$}
\address{(1) Istituto dei Sistemi Complessi, CNR, C.P. 10, 
Monterotondo Stazione, Roma I-00016}
\address{(2) Institut des NanoSciences de Paris, CNRS, Universit\'{e} 
Pierre et Marie Curie, 140 rue de Lourmel, 75015 Paris}
\date{\today }

\begin{abstract}
   The exciton-exciton effective scattering which rules the time 
evolution of two excitons is studied as a function of initial 
momentum difference, scattering angle and electron-to-hole mass 
ratio. We show that  this effective scattering can collapse for 
energy-conserving configurations provided that the difference between 
the two initial exciton momenta is larger than a threshold value. 
Sizeable scatterings then exist in the forward direction only. We 
even find that, for an electron-to-hole mass ratio close to 1/2, the 
exciton-exciton effective scattering stays close to zero in all 
directions when the difference between the initial exciton momenta 
has a very specific value. This unexpected  but quite remarkable 
collapse comes from tricky compensation between direct and exchange 
Coulomb processes which originates from the fundamental 
undistinguishability of the exciton fermionic components.
\end{abstract}

\maketitle

\section{Introduction}

It is now commonly accepted that the composite boson nature of excitons plays 
a key role in their many-body physics: The undistinguishability of 
the two carriers from which excitons are constructed leads to all 
kinds of elaborate exchange processes which can hardly be handled 
within a na\"{\i}ve bosonic framework.

These exchange processes enter exciton-exciton interactions which are known to be the main source of nonlinearity in optical properties of semiconductors. Among them, we can cite stimulated scattering\cite{Savvidis}, polarization change\cite{Krizhanovskii}, bistable\cite{Baas} and multistable\cite{Gippius} behaviors and exciton spin relaxation\cite{Garro}.

Exchange between indistinguishable quantum particles has also been shown to play a fundamental role in the physics of Bose-Einstein condensation.\cite{nozieres} Interest of
this argument is renewed by recent observations
of polariton condensates\cite{Kasprzak,balili}, polaritons being mixed states of one exciton and one photon. In pioneering experiments\cite{amo} addressed to fluid propagation of a coherent polariton gas\cite{Malpuech,Utsunomiya}, the formation of vortices\cite{Richard} has just been evidenced. It is however worth noting that, although theoretically predicted much longer ago\cite{Moskalenko,Keldysh}, the experimental observation of Bose-Einstein condensation of a pure exciton gas remains a challenge \cite{Snoke,Voros}, in spite of very many different attempts \cite{Wouters,Ciuti,Inoue,Shumway,Schindler}.  Possible reason for not observing exciton Bose-Einstein condensation can actually be due to the fact that excitons must condense into a dark state \cite{dark}. Here too, carrier exchanges play a key role since exchange between two opposite spin bright excitons produce two opposite spin dark excitons.

A correct handling of the exciton composite nature thus is a request, not only from the theoretical point of view, but also to correctly understand the experimental data.

One of the most drastic mathematical difference between elementary 
bosons and composite bosons made of two free fermions with momenta 
$(\v k_e,\v k_h)$, as the Wannier excitons, is the fact that the 
prefactor in the $N$-particle closure relation is $(1/N!)^2$ when the 
particle composite nature is kept while it is $(1/N!)$ only when these 
particles are replaced by elementary bosons \cite{closure relation 
1,closure relation 2}. This prefactor difference proves that the 
formal replacement of Wannier excitons by elementary bosons is a 
dream, even in the extreme dilute limit of just $N=2$ excitons, 
because all sum rules which result from closure relation, are going 
to be different, whatever the exciton-exciton effective scatterings.

By contrast, it is worth noting that the closure relation 
for Frenkel excitons \cite{closure relation 
2,Frenkel1,Frenkel2,Frenkel3}, which are made of electron-hole pairs 
localized on the same ion site, has the same $(1/N!)$ prefactor as 
the one of elementary bosons. The reason is that, instead of two 
degrees of freedom $(\v k_e,\v k_h)$, Frenkel excitons have one only: 
the ion site $n$.

 From a mathematical point of view, the composite nature of particles 
constructed on two free fermions makes the exciton basis for $N$-pair 
states overcomplete --- except for $N=1$. As a mere consequence, the 
$N$-Wannier-exciton states are not orthogonal. While a non-orthogonal 
basis is rather easy to handle, the intrinsic overcompleteness of the
Wannier-exciton state basis cannot be eliminated in a self-consistent way. 
This makes all attempts \cite{Laikhtman,Rombouts}
to work with an orthogonalized exciton state set, 
doomed to failure because the difficulty is not so much to find a 
procedure to orthogonalize the states but to reduce their number 
consistently. In the following, we will restrict to Wannier excitons 
since those are the ones for which the composite-boson nature shows up 
the most dramatically.

Having, on the one hand, understood the intrinsic difficulty linked 
to the overcompleteness of $N$-pair states when written in terms of 
exciton operators, being, on the other hand, fully convinced that 
these exciton states constitute the relevant basis 
\cite{Mukamel1,Mukamel2,Mukamel3}
for a proper description of many-body effects in a dilute system of 
electron-hole pairs, we have recently constructed a formalism 
\cite{Physics Report} which allows us to handle this overcompleteness 
in an exact way. The conceptual difference between our formalism and 
the Green function formalism developed long ago for elementary quantum 
particles lies in the fact that this composite boson formalism uses 
an operator algebra based on commutators between exciton 
creation operators \cite{Kali}, while the Green function formalism 
relies on scalars only.

The link between the overcompleteness of exciton states and the 
exciton composite nature is evidenced through the relation
\begin{equation}
B_i^\dag B_j^\dag=-\sum_{mn}\lambda\left(^{n\ \,j}_{m\
i}\right)\,B_m^\dag B_n^\dag\ .
\end{equation}
which comes from the two different ways to associate two electrons 
and two holes into two Wannier excitons. $B_i^\dag$  is the creation 
operator for exciton $i$ having $\v K_i$ as center-of-mass momentum 
and $\nu_i$ as relative motion index. The 2 by 2 Pauli scattering 
$\lambda\left(^{n\ \,j}_{m\ i}\right)$, shown in Fig.1, describes 
fermion exchange between two excitons in states $(i,j)$ in the absence of 
fermion interaction. The real difficulty with composite excitons is 
to produce a formalism unchanged with respect to the above identity. 
Our composite-boson many-body theory \cite{Physics Report} does it in 
an exact self-consistent way.
\begin{figure}[t]
\includegraphics[scale=0.35]{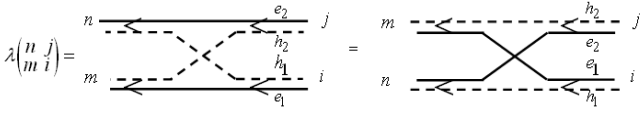}
\caption{Pauli scattering $\lambda\left(^{n\ \,j}_{m\
i}\right)$ for fermion exchange between two excitons starting in "in" 
states $(i,j)$ and ending in "out" states $(m,n)$. Electrons are represented by solid lines and holes by dashed lines.}
\end{figure}

As $\lambda\left(^{n\ \,j}_{m\
i}\right)$ in Eq.(1) is a dimensionless quantity, such pure exchange 
scatterings cannot appear alone in effective scatterings ruling the 
time evolution of two excitons because these effective scatterings 
must be energy-like quantities. However, when mixed with Coulomb 
process, these carrier exchanges, which come from the intrinsic 
undistinguishablity of the exciton fermionic components, become 
crucial because they readily lead to \emph{six} different energy-like 
scatterings between two excitons. These depend on how the carriers of 
the two excitons are associated in the ``in'' and ``out'' states 
$(i,j)$ and $(m,n)$. The resulting six different scatterings are 
shown in Figs.(2,3).
In  the direct Coulomb scattering $\xi^\mathrm{dir}\left(^{n\ j}_{m\ 
i}\right)$, shown in Fig.2(a), the
excitons $m$ and $i$ are made with the same carriers while in $\xi^\mathrm{dir}
\left(^{m\ j}_{n\ \,i}\right)$ obtained by a $(m,n)$ permutation, 
their two carriers are different, so that $\xi^\mathrm{dir}\left(^{m\ 
j}_{n\ \,i}\right)$ can be seen as a direct Coulomb scattering 
followed by two Pauli scatterings for carrier exchange in the absence 
of carrier interaction [see Fig.2(b)].
\begin{figure}[t]
\includegraphics[scale=0.4]{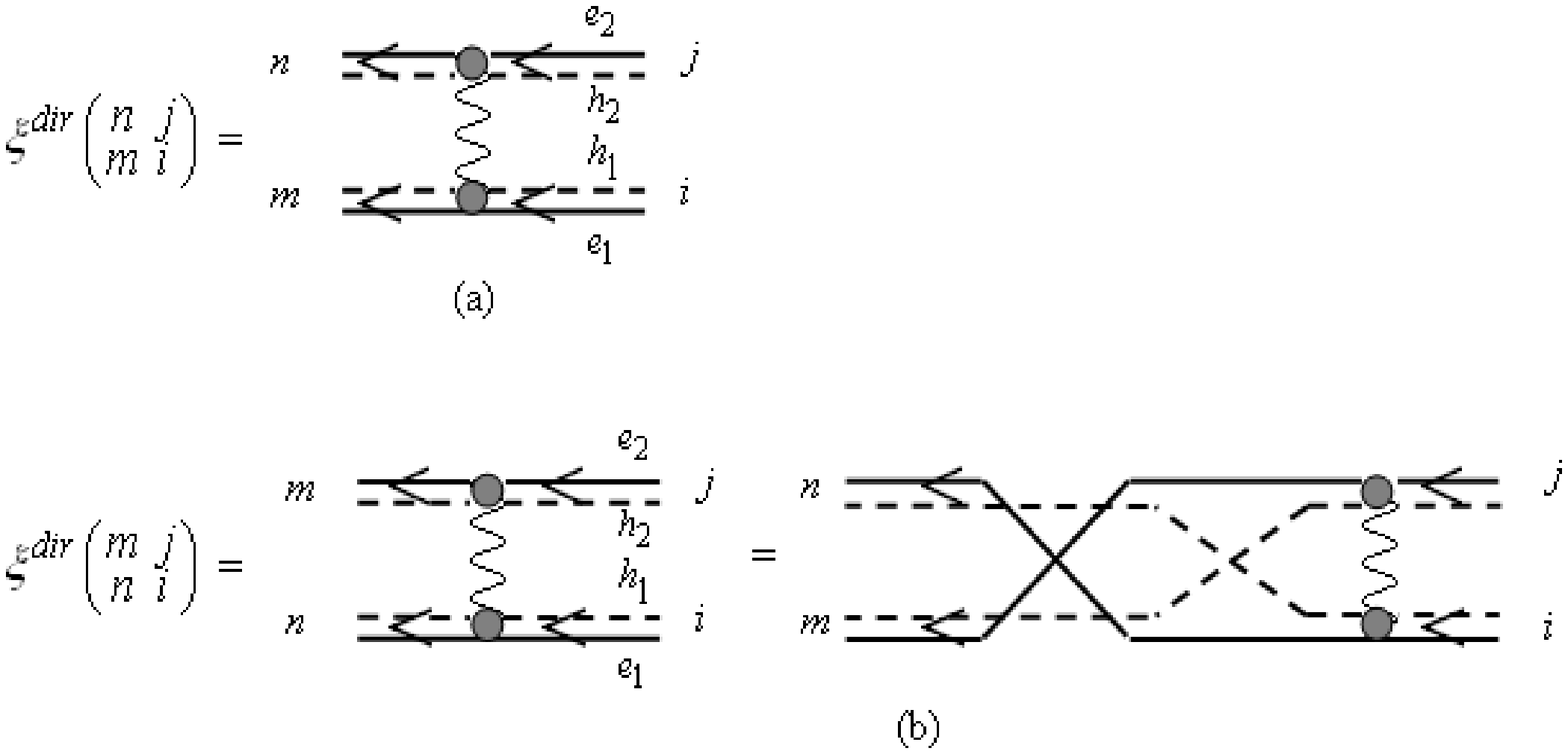}
\caption{(a): Direct interaction scatterings  between two excitons 
starting in "in" states $(i,j)$ and ending in "out" states $(m,n)$ for 
carrier interaction in the absence of carrier exchange. (b): A $(m,n)$
permutation can be seen as the result of a double carrier exchange.}
\end{figure}
The direct Coulomb scattering $\xi^\mathrm{dir}\left(^{n\ j}_{m\ 
i}\right)$ is given by \cite{Physics Report}
\begin{eqnarray}
\xi^\mathrm{dir}\left(^{n\ \,j}_{m\ i}\right)=\int\{d\v 
r\}(V_{e_1e_2}+V_{h_1h_2}-V_{e_1h_2}-V_{e_2h_1})\nonumber\\
\times\langle m|\v r_{e_1},\v r_{h_1}\rangle\langle n|\v r_{e_2},\v 
r_{h_2}\rangle
\langle\v r_{h_2},\v r_{e_2}|j\rangle\langle\v r_{h_1},\v r_{e_1}|i\rangle.
\end{eqnarray}
$\langle\v r_{h_1},\v r_{e_1}|i\rangle$ is the wave function of an exciton 
in state $i$, its electron being located at $\v r_{e_1}$ and its hole 
at $\v r_{h_1}$. Coulomb interaction between two electrons reads 
$V_{e_1e_2}=e^2/|\v r_{e_1}-\v r_{e_2}|$ and similarly for the other 
Coulomb terms. The above expression of $\xi^\mathrm{dir}\left(^{n\ j}_{m\ 
i}\right)$ visually follows from the 
diagrammatic representation of Fig.2(a).

In addition to direct Coulomb scatterings, two excitons can also have 
exchange Coulomb scatterings.
In   $\xi^\mathrm{in}\left(^{n\ \,j}_{m\ i}\right)$ and
$\xi^\mathrm{out}\left(^{n\ \,j}_{m\ i}\right)$, the excitons $m$ and 
$i$ have the same electron but a different hole, while in 
$\xi^\mathrm{in}\left(^{m\
j}_{n\ \,i}\right)$ and
$\xi^\mathrm{out}\left(^{m\
j}_{n\ \,i}\right)$ obtained from a $(m,n)$ permutation, they have 
the same hole but a different electron. These exchange Coulomb 
scatterings result  from a succession of direct Coulomb scattering 
and Pauli scattering for carrier exchange. Fig.3(a) shows that
$\xi^\mathrm{in}\left(^{n\ \,j}_{m\ i}\right)$ reads as 
$\xi^\mathrm{dir}\left(^{n\ \,j}_{m\ i}\right)$ with $(\v r_{h_1},\v 
r_{h_2})$ exchanged in the $(m,n)$ wave functions.

\begin{figure*}[t]
\includegraphics[scale=0.65]{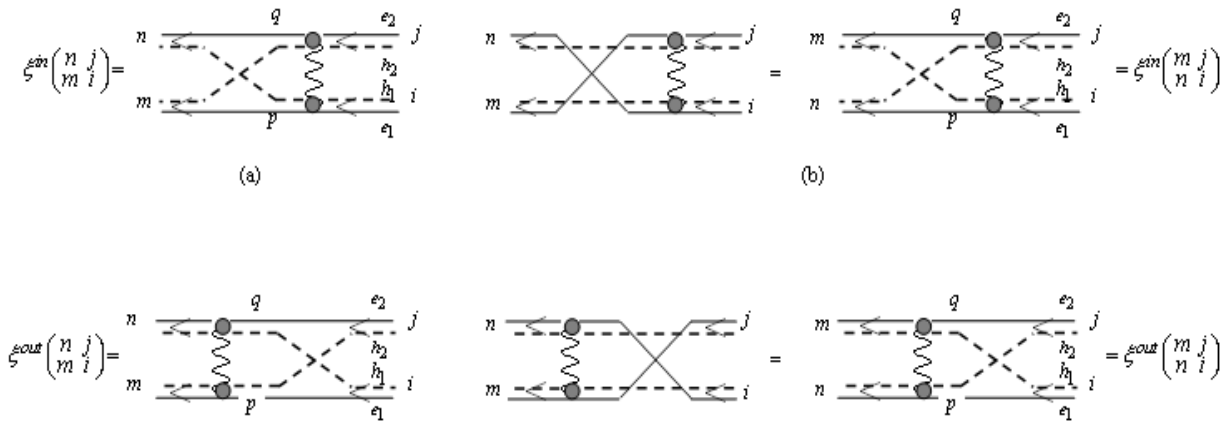}
\caption{Coulomb exchange scatterings: in  $\xi^\mathrm{in}$, the 
Coulomb processes take place between the "in" excitons, while in 
$\xi^\mathrm{out}$ they take place between the "out" excitons.  These 
exchange Coulomb scatterings can be seen as a succession of a direct 
Coulomb scattering and a Pauli scattering for carrier exchange in the 
absence of carrier interaction.}
\end{figure*}

It turns out that, as briefly rederived below, $\xi^\mathrm{out}$ in 
which Coulomb interaction takes place  \textit{after} carrier 
exchange, does not enter the effective scattering ruling the time 
evolution of two excitons  \cite{PRL,R.C.}. This is due to a very 
fundamental reason linked to symmetry breaking in the evolution 
towards positive time. It however is of interest to note that, as 
physically expected, a symmetry between these two exchange Coulomb 
scatterings exists in the large time limit: indeed, we do have 
\cite{Physics Report}
\begin{equation}
\xi^\mathrm{in}\left(^{n\ \,
j}_{m\ i}\right)-\xi^\mathrm{out}\left(^{n\ \,
j}_{m\ i}\right)=(E_m+E_n-E_i-E_j)\,\lambda\left(^{n\ \,
j}_{m\ i}\right)\ .
\end{equation}
where $E_i$ is the $i$ exciton energy, so that the two exchange 
Coulomb scatterings are equal for energy-conserving processes, i.e., 
when time reversal is  expected.

Even if $\xi^\mathrm{out}$ does not enter the effective scattering 
ruling the time evolution of two excitons, we are nevertheless left 
with four different energy-like quantities. Being equally relevant 
since they only differ by the intrinsic fermion undistinguishability, 
these four scatterings must appear on equal footing in the effective 
scattering ruling the time evolution of two excitons. Consequently, 
this effective scattering must read as a linear combination of four 
terms, each of these four terms containing two Coulomb attractions and two Coulomb repulsions. Such a complex structure is a direct consequence of the 
particle composite nature. Some tricky compensations can then take 
place in this linear combination, to possibly end with an effective 
scattering, either very close to zero, or even exactly equal to zero 
for some specific configurations; the corresponding initial state 
can then be seen as ``frozen'' at first order in Coulomb processes.

In a previous work \cite{R.C.}, we found that the effective 
scattering of two excitons having \emph{same} initial momentum cancels for a 
finite value of the momentum transfer. This particular transfer 
however has no physical relevance because it does not correspond to 
process in which energy is conserved. Being still puzzled by this 
somewhat unexpected cancellation, we wanted to reconsider the problem 
more in details in order to see if the exciton-exciton effective scattering 
which rules the time evolution of two excitons can cancel for some
energy-conserving configurations. This is the purpose of the present work.

We here show that, indeed, there are some configurations in which the 
effective scattering ruling the time evolution of two excitons does 
cancel while energy is conserved. As a result, the corresponding 
scattering configurations are forbidden at first order in the 
interaction. For some electron-to-hole mass ratio close to 1/2, the 
effective exciton-exciton scattering can even stay very close to zero 
in all directions provided that the initial momentum difference has a 
very specific value. This particular initial configuration then 
appears as somewhat magic because excitons do not 
scatter through first-order Coulomb process. Such a cancellation 
however requires initial exciton momenta above a threshold value 
which is far larger than the typical photon momenta, i.e., the momenta of the
photocreated excitons. Excitons in their relative motion ground state with a larger kinetic energy can however be formed through collisions between excited state excitons resulting from photon excitation above the absorption edge. However, independently from its possible observation, it is of importance to understand that the exciton-exciton effective scattering can collapse as a result of the exciton composite nature. To reveal its existence thus constitutes a relevant part of the overall understanding of exciton-exciton interaction.

The present paper is organized as follows.

In section II, we construct the effective scattering which rules the 
time evolution of two excitons as imposed by the particle quantum 
nature. In order to better grasp the importance of the fermion/boson 
nature of the particles as well as the consequence of fermionic 
components in this effective scattering, we here briefly rederive the 
time evolution of two elementary fermions, two elementary bosons, and 
two Wannier excitons. This allows us to evidence that a 
possible collapse of the effective scattering is a fundamental property of 
elementary fermions, this collapse appearing above a threshold only 
in the case of composite bosons made of two fermions.

In section III, we study the possible cancellation of this effective 
exciton-exciton scattering for energy conserving configurations. To 
this end, we perform a numerical calculation of the ``in'' exchange 
Coulomb scattering appearing in this effective scattering in the most 
general case, i.e.,  for initial excitons having different momenta 
and arbitrary mass ratio. We then restrict to energy-conserving configurations and study the dependence of the 
effective scattering on exciton momentum difference, scattering angle 
and electron-to-hole mass ratio. We pay particular attention to the 
magic configuration in which the effective scattering stays close to 
zero in all directions.

In section IV, we conclude.

\section{Effective scattering for the time evolution of two quantum particles}

\subsection{Relevant coordinates}

In order to analyse the exciton momentum configuration possibly 
leading to a cancellation of the exciton effective scattering, we will, for simplicity, restrict to 2D scatterings in which 
the excitons stay in their relative motion ground state, i.e., 
processes in which all the relative motion indices $\nu$ are equal to $\nu_0$. This restriction requires
two conditions to be met:

1) The quantum well should be narrow enough to possibly consider one confined  level only.

2) The exciton momenta, P, should be small enough to avoid scattering towards unbound electron-hole pairs. This essentially imposes an exciton kinetic energy smaller than the 2D binding energy, namely, $P^2/2M\leq 4/2\mu a_X^2$, where $M=m_e+m_h$ is the exciton center-of-mass mass, $\mu^{-1}=m_e^{-1}+m_h^{-1}$ the inverse exciton relative motion mass and $a_X$ the 3D exciton Bohr radius. For $\tilde{P}=a_XP$, this condition reads $\tilde{P}\leq 2(1+\alpha)/\sqrt{\alpha}$. It is shown in Fig.4 as a function of the mass ratio $\alpha=m_e/m_h.$ 
\begin{figure}[t]
\includegraphics[scale=0.35]{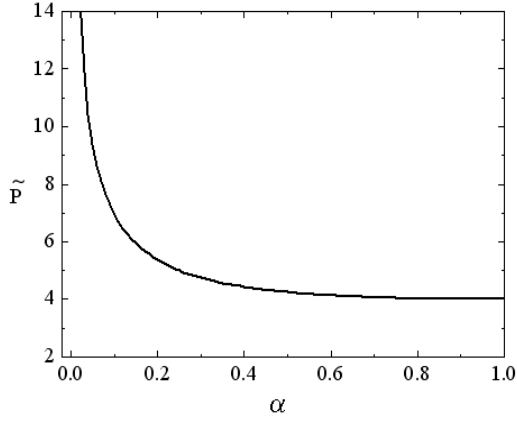}
\caption{Exciton momenta in 3D Bohr radius unit $\tilde{P}=a_XP$, for which the exciton kinetic energy is equal to its binding energy, as a function of mass ratio $\alpha=\frac{m_e}{m_h}$.}
\end{figure}

Also for 
simplicity, we will not here consider the exciton spin degrees of 
freedom: this physically corresponds to take the electrons (holes) of 
the two excitons with same spin.
\begin{figure}[t]
\includegraphics[scale=0.35]{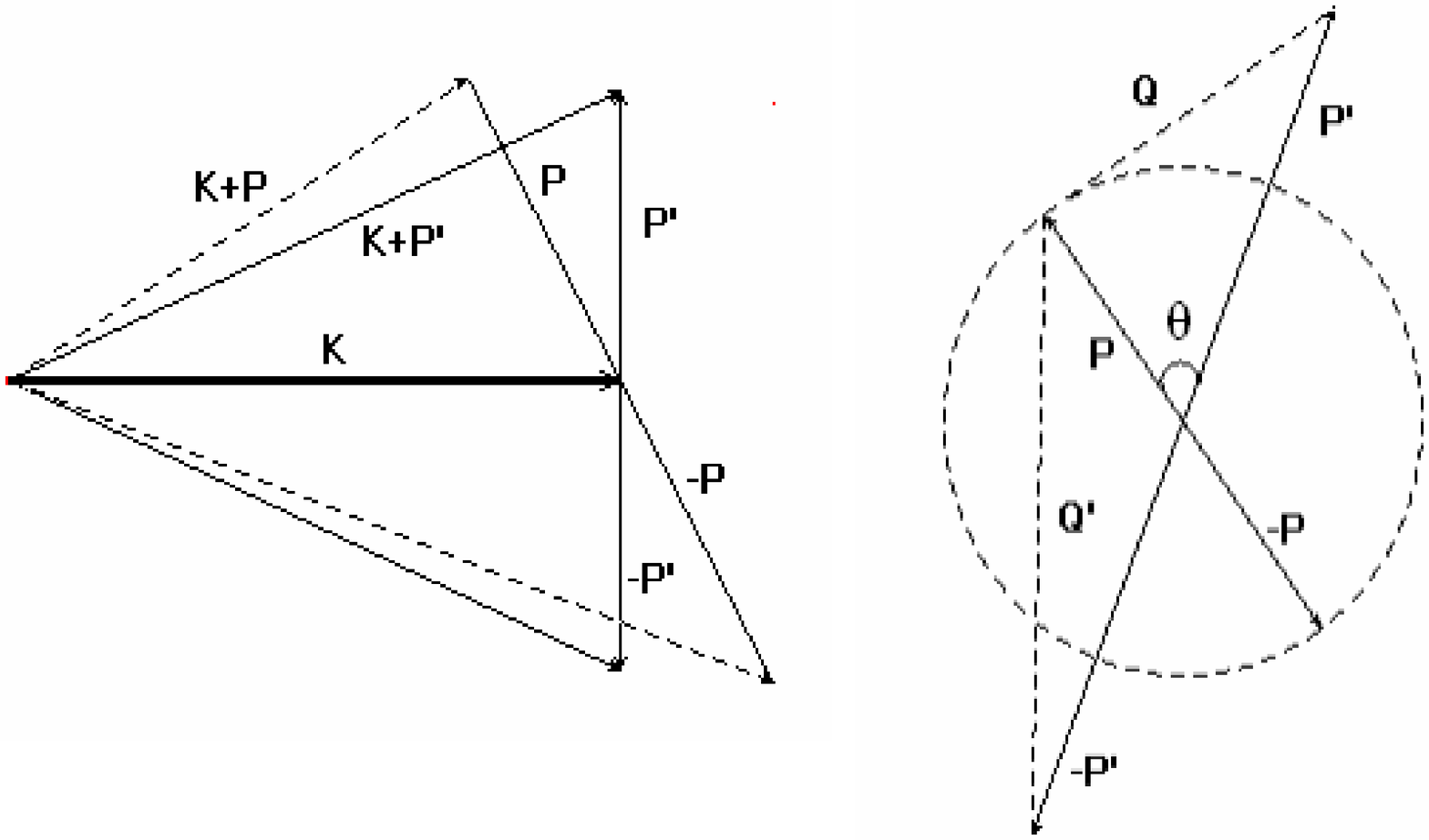}
\caption{(a): In the laboratory frame, the "in" 
excitons with momenta $\v K_i=\v K+\v P$ and $\v K_j=\v K-\v P$ 
transform into "out" excitons with momenta $\v K_m=\v K+\v P'$ and 
$\v K_n=\v K-\v P'$. (b): In the center-of-mass frame, which 
corresponds to set $\v K=0$, these excitons have momenta 
$(\v P,-\v P)$ and $(\v P',-\v P')$. Energy conserving processes, 
which are the relevant ones in the large time limit, lead to $P=P'$: momenta follow
a simple rotation in the center-of-mass frame, as 
shown by the dashed circle.}
\end{figure}

Since the total center-of-mass momentum $2\v K$ of two 
excitons is conserved in a scattering process, we are led to 
write the center-of-mass momenta of the two initial excitons as 
$\v K_i=\v K+\v P$ and $\v K_j=\v K-\v P$, while the ones of the two 
final excitons are written as $\v K_m=\v K+\v P'$ and $\v K_n=\v K-\v P'$ [see 
Fig.5(a)]. As physical results 
cannot depend on frame momentum, we can, without any loss of 
generality, set $\v K$ equal to zero. This leads us to rewrite the 
effective scattering for the configuration of interest as, [see 
Fig.6],
\begin{equation}
\xi^\mathrm{eff}\left(^{n\ \,j}_{m\ i}\right)\equiv 
\xi^\mathrm{eff}\left(^{-\v P'\ -\v P}_{\ \,\,\v P'\ \ \ \v P}\right) 
\ .
\end{equation}

It will also appear as convenient
to introduce the two momentum transfers $(\v Q,\v Q')$ of this 
scattering process. These are defined as
\begin{equation}
\v P+\v Q=\v P' \hspace{2cm} \v P+\v Q'=-\v P'\ :
\end{equation}
due to carrier undistinguishability, exciton with 
initial momentum $\v P$ can as well end with the final momentum $\v P'$ or 
$-\v P'$,  which 
corresponds to change $\v Q$ into $\v Q'$ [see Fig.5(b)].
\begin{figure}[b]
\includegraphics[scale=0.3]{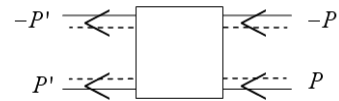}
\caption{Scattering of two excitons with initial momenta $(\v P,-\v P)$ and 
final momenta $(\v P',-\v P')$ in the center-of-mass frame.}
\end{figure}

Since $\v Q=\v P'-\v P$, while $\v Q'=-\v P'-\v P$, these momentum 
transfers are such that $Q=Q'$ for $\v P.\,\v P'=0$. We can also note 
that, for scatterings staying within the same exciton relative motion 
subspace, i.e., for excitons all having the same relative motion index 
$\nu_0$, energy conservation imposes $(\v K+\v P)^2+(\v K-\v P)^2=(\v 
K+\v P')^2+(\v K-\v P')^2$, i.e., $P=P'$: the scattered momentum then
evolves on a radius $P$ circle. This $P=P'$ condition also reads 
$\v Q.\v Q'=0$. As $\v Q+\v Q'=-2\v P$, the condition $\v Q.\v 
Q'=0$ implies $Q^2+Q'^2=4P^2$. This shows that, for a given $P$, the 
final states having the energy of the initial state, are fully 
determined by the scattering angle $\theta$ between $\v P$ and $\v 
P'$.

As we are mainly interested in energy conserving processes, we will ultimately 
study the effective scattering $\xi^\mathrm{eff}\left(^{-\v 
P'\ -\v P}_{\ \,\,\v P'\ \ \ \v P}\right)$ of two excitons as a 
function of the scattering angle $\theta$ and half the initial 
exciton momentum difference $P=|\v K_i-\v K_j|/2$, for various electron-to-hole mass 
ratios.

\subsection{Effective scattering for elementary particles}

To better grasp the importance of the particle composite nature and 
to relate the possible collapse of the effective $2\times 2$ 
scattering to the particle fermionic/bosonic nature, let us first consider two 
\emph{elementary} quantum particles having initial momenta $(\v P,-\v 
P)$ and final momenta $(\v P',-\v P')$ in the center-of-mass frame 
$(\v K=\v 0$).
The time evolution of the initial state $(\v P,-\v P)$ is given by
\begin{equation}
|\psi_t\rangle=e^{-iHt}C_{\v P}^\dag C_{-\v P}^\dag|v\rangle\ ,
\end{equation}
where $C_{\v P}^\dag$ creates the elementary particle of interest 
with momentum $\v P$. Depending on the particle quantum nature, these 
operators are such that
\begin{equation}
C_{\v P_1}C_{\v P_2}^\dag+\eta C_{\v P_2}^\dag C_{\v P_1}=\delta_{\v 
P_2,\v P_1}\ ,
\end{equation}
with $\eta =1$ for fermions and $\eta=-1$ for bosons. The above equation leads to
$\langle v|C_{-\v P'}C_{\v P'}C_{\v P}^\dag C_{-\v 
P}^\dag|v\rangle=\delta_{\v P',\v P}+\eta\delta_{\v P',-\v P}$.

The simplest way to calculate $|\psi_t\rangle$ is to use the integral 
representation of the exponential. For $t>0$, it reads
\begin{equation}
e^{-iHt}=\int_{-\infty}^{+\infty}\frac{dx}{(-2i\pi)}\,\frac{e^{-it(x+iO_+)}}{x+iO_+-H}\ 
,
\end{equation}
where $O_+$ is an arbitrary positive constant. The probability to go 
from $(\v P,-\v P)$ to $(\v P',-\v P')$ is thus given by
\begin{eqnarray}
\langle v|C_{-\v P'}C_{\v P'}e^{-iHt}C_{\v P}^\dag C_{-\v 
P}^\dag|v\rangle=\int_{-\infty}^{+\infty}\frac{dx}{(-2i\pi)}e^{-it(x+iO_+)}\nonumber\\
\times\langle v|C_{-\v P'}C_{\v P'}\frac{1}{x+iO_+-H}C_{\v P}^\dag 
C_{-\v P}^\dag|v\rangle\ .\hspace{1cm}
\end{eqnarray}
To go further and calculate the matrix element in this integral, we 
use the following identity valid for $H=H_0+V$, this identity being 
the key for correlation effects with elementary quantum particles,
\begin{equation}
\frac{1}{z-H}=\frac{1}{z-H_0}+\frac{1}{z-H}\, V\, \frac{1}{z-H_0}\ .
\end{equation}

For $H_0=\sum_{\v k}\epsilon_{\v k}C_{\v k}^\dag C_{\v k}$ and $V$ given by
\begin{equation}
V=\frac{1}{2}\sum_{\v q\neq\v 0}V_{\v q}\sum_{\v k_1,\v k_2}C_{\v 
k_1+\v q}^\dag C_{\v k_2-\v q}^\dag
C_{\v k_2} C_{\v k_1}\ ,
\end{equation}
this leads us to write, at first order in the interaction,
\begin{eqnarray}
\langle v|C_{-\v P'}C_{\v P'}e^{-iHt}C_{\v P}^\dag C_{-\v 
P}^\dag|v\rangle\simeq\hspace{3cm}\nonumber\\
e^{-2i\epsilon_{\v P}t}(\delta_{\v P,\v P'}+\eta\delta_{\v P,-\v 
P'})\hspace{3cm}\nonumber\\
+(V_{\v P'-\v P}+\eta V_{\v P'+\v P})\frac{e^{-2i\epsilon_{\v 
P'}t}-e^{-2i\epsilon_{\v P}t}}{2\epsilon_{\v P'}
-2\epsilon_{\v P}}+\cdots
\end{eqnarray}
This shows that, due to the quantum particle undistinguishability, 
the effective scattering which rules the time evolution of two 
elementary particles starting in state $(\v P,-\v P)$ and ending in a 
different state $(\v P',-\v P')$ is made of two processes which 
differ by a particle exchange,
\begin{equation}
\xi^\mathrm{eff}\left(^{-\v P'\ -\v P}_{\ \,\,\v P'\ \ \ \v 
P}\right)=V_{\v P'-\v P}+\eta V_{\v P'+\v P}=V_{\v Q}+\eta V_{\v Q'}\ 
.
\end{equation}

In the case of two fermions, this effective scattering reduces to 
$V_{\v Q}-V_{\v Q'}$, the minus sign being standard for process 
associated to fermion exchange. This shows that the effective 
scattering ruling the time evolution of two elementary fermions 
cancels for $Q=Q'$, i.e., for $\v P'$ perpendicular to $\v P$, with in addition
$P'=P$ in the case of energy-conserving processes. Let us 
stress that this effective scattering sign change should not be taken 
as a change from repulsion to attraction between the particles at 
hand: this latter characteristic is fully linked to the sign of the 
elementary scattering $V_{\v Q}$ in the Hamiltonian, not to the sign of the effective 
scattering ruling the time evolution of two fermions, as obvious from 
the fact that this effective scattering anyway appears as a square 
modulus in the Fermi golden rule.

By contrast, the effective scattering of two elementary bosons is 
given by $V_{\v Q}+V_{\v Q'}$, so that such a cancellation does not 
occur. The possible cancellation of the effective scattering ruling 
the time evolution of two elementary quantum particles thus appears 
as a characteristic of the particle fermionic nature. A memory of 
this fermionic cancellation is going to show up in the case of 
excitons made of two free fermions, but above a momentum threshold only.

\subsection{Effective scattering for two-fermion particles}

We now turn to the time evolution of two Wannier excitons made of linear combination of free fermion pairs. Let 
$B_i^\dag$ be the creation operator of one exciton in state $i$. This 
operator is such that $(H-E_i)B_i^\dag|v\rangle=0$. Due to the 
exciton composite nature, the scalar product of two-exciton states is 
given by \cite{Physics Report}
\begin{equation}
\langle 
\phi_{mn}|\phi_{ij}\rangle=\left[\delta_{m,i}\delta_{n,j}-\lambda\left(^{n\ 
\,j}_{m\ i}\right)\right]+
[m\leftrightarrow n]\ ,
\end{equation}
where $|\phi_{ij}\rangle=B_i^\dag B_j^\dag|v\rangle$ and 
$\lambda\left(^{n\ \,j}_{m\ i}\right)$ is the Pauli scattering for 
fermion exchange in the absence of fermion interaction, shown in 
Fig.1.

To get the time evolution of the two-exciton state $B_i^\dag 
B_j^\dag|v\rangle$, we use a procedure similar to the one we have 
used in the case of elementary quantum particles. Equation (8) allows 
us to write
\begin{eqnarray}
\langle\phi_{mn}|e^{-iHt}|\phi_{ij}\rangle=\int_{-\infty}^{+\infty}\frac{dx}{(-2i\pi)}\,e^{-it(x+iO_+)}\nonumber\\
\times\langle\phi_{mn}|\frac{1}{x+iO_+-H}|\phi_{ij}\rangle\ .
\end{eqnarray}
Since the semiconductor Hamiltonian does not split in terms of 
exciton operators as $H_\mathrm{X}+V_\mathrm{XX}$, we cannot use 
Eq.(10). Correlations betwen excitons then follow from a 
similar equation in which enters the exciton creation operator, namely \cite{Physics Report},
\begin{equation}
\frac{1}{z-H}B_i^\dag=B_i^\dag\frac{1}{z-H-E_i}+\frac{1}{z-H}\,V_i^\dag\frac{1}{z-H-E_i}\ 
,
\end{equation}
where the operator $V_i^\dag=[H,B_i^\dag]-E_iB_i^\dag$ describes the 
interactions of exciton $i$ with the rest of the system. To go further, we 
introduce the direct Coulomb scattering
$\xi^\mathrm{dir}\left(^{n\ \,j}_{m\ i}\right)$ formally defined as 
\cite{Physics Report}
\begin{equation}
[V_i^\dag,B_j^\dag]=\sum_{mn}\xi^\mathrm{dir}\left(^{n\ \,j}_{m\ 
i}\right)B_m^\dag B_n^\dag\ .
\end{equation}
Its precise value is given in Eq.(2) and its diagrammatic 
representation is shown in Fig.2(a).

It is then easy to show that, to lowest order in the interaction,
\begin{eqnarray}
\langle\phi_{mn}|\frac{1}{z-H}|\phi_{ij}\rangle\simeq 
\frac{1}{z-E_{ij}}\langle\phi_{mn}|\phi_{ij}\rangle\hspace{2cm}\nonumber\\
+\frac{1}{(z-E_{mn})(z-E_{ij})}\sum_{p,q}\langle\phi_{mn}|\phi_{pq}\rangle\xi^\mathrm{dir}\left(^{q\ 
j}_{p\ i}\right)\ ,
\end{eqnarray}
where $E_{ij}=E_i+E_j$. Using the scalar product of two-exciton 
states given in Eq.(14), we end with
\begin{eqnarray}
\langle\phi_{mn}|e^{-iHt}|\phi_{ij}\rangle\simeq \hspace{4cm}\nonumber\\
e^{-iE_{ij}t}\left\{\left[\delta_{m,i}\delta_{n,j}-
\lambda\left(^{n\ \,j}_{m\ i}\right)\right]+[m\leftrightarrow 
n]\right\}\nonumber\\
+\frac{e^{-iE_{mn}t}-e^{-iE_{ij}t}}{E_{mn}-E_{ij}}\,\xi^\mathrm{eff}\left(^{n\ 
\,j}_{m\ i}\right),
\end{eqnarray}
where the effective scattering ruling the time evolution of the two 
excitons $(i,j)$ is given by
\begin{equation}
\xi^\mathrm{eff}\left(^{n\ \,j}_{m\ 
i}\right)=\left[\xi^\mathrm{dir}\left(^{n\ \,j}_{m\ 
i}\right)-\xi^\mathrm{in}\left(^{n\ \,j}_{m\ 
i}\right)\right]+[m\leftrightarrow n]\ ,
\end{equation}
with $\xi^\mathrm{in}\left(^{n\ \,j}_{m\ i}\right)$ being the ``in'' 
exchange Coulomb scattering defined as
\begin{equation}
\xi^\mathrm{in}\left(^{n\ \,j}_{m\ 
i}\right)=\sum_{p,q}\lambda\left(^{n\ \,q}_{m\ 
p}\right)\xi^\mathrm{dir}\left(^{q\ j}_{p\ i}\right)\ ,
\end{equation}
and shown in Fig.3(a).

The undistinguishability of the electron-hole components of the 
excitons leads to an effective scattering made of four terms instead 
of two as in the case of elementary quantum particles: starting from 
the direct Coulomb scattering $\xi^\mathrm{dir}\left(^{-\v P'\ -\v 
P}_{\ \,\,\v P'\ \ \ \v P}\right)$, the three other scatterings 
correspond either to exchange one electron or one hole as in the two 
exchange Coulomb scatterings $\xi^\mathrm{in}$ of Figs.3(a,b), or to 
exchange the two carriers, which is nothing but a $(\v 
P'\leftrightarrow -\v P')$ exchange in $\xi^\mathrm{dir}$ as shown in 
Fig.2(b). Since a fermion exchange brings a minus sign, we end with 
an effective scattering for the process of Fig.6 made of two terms 
with a plus sign and two terms with a minus sign:
\begin{eqnarray}
\xi^\mathrm{eff}\left(^{-\v P'\ -\v P}_{\ \,\,\v P'\ \ \ \v 
P}\right)=\xi^\mathrm{dir}\left(^{-\v P'\ -\v P}_{\ \,\,\v P'\ \ \ \v 
P}\right)-\xi^\mathrm{in}\left(^{-\v P'\ -\v P}_{\ \,\,\v P'\ \ \ \v 
P}\right)\nonumber\\
+(\v P'\leftrightarrow -\v P')\ ,
\end{eqnarray}
each of these four terms being actually made of two Coulomb 
repulsions and two Coulomb attractions, as seen from Eq.(2). Due to 
such a complex structure, it is far from obvious to physically guess 
the sign of the resulting effective scattering and its possible 
cancellation.

We are going to show that, as for two elementary fermions, the 
effective scattering for the time evolution of two excitons can 
cancel. However, for energy conserving process, this cancellation 
requires a difference between initial momenta larger than a threshold 
value: for lower initial momentum difference, the exciton bosonic nature dominates: the 
effective scattering, like for two elementary bosons, keeps a constant sign 
for all scattering configurations.

Actually, due to a quite subtle interplay between the various Coulomb 
contributions existing in this effective scattering, interplay which deeply depends on possible symmetry between electron and hole, we even find 
that, when the electron-to-hole mass ratio is close to 1/2, the 
effective scattering stays essentially equal to zero in all 
directions, provided that the initial momentum 
difference has a very specific value. While the cancellation of the effective scattering for a 
particular value of the scattering angle is quite standard for elementary 
fermions, the possible cancellation of this effective scattering \emph{in 
all scattered directions} is far more subtle, being deeply linked to 
the exciton composite nature. It is worth noting that the mass ratio 1/2 to have this somewhat magic cancellation essentially separates hydrogen-like excitons $(m_e\ll m_h)$ from positronium-like excitons for which the electron and hole play a quite symmetrical role.

\subsection{Various contributions to the effective scatterings}

As seen from Eqs.(20) or (22), the effective scattering ruling the 
time evolution of two excitons is made of two direct terms and two 
exchange terms.

\subsubsection{Direct terms}

In previous works \cite{OBM,R.C.}, we showed that the direct exciton-exciton
scattering $\xi^\mathrm{dir}\left(^{-\v P'\ -\v P}_{\ \,\,\v P'\ \ \ 
\v P}\right)$ in which excitons with momenta $\v P$ and $\v P'$ are 
made with the same electron-hole pair, can be written analytically in 
terms of the exciton momentum transfer $\v Q=\v P'-\v P$. In the case 
of 2D ground state excitons for which $\langle\v 
r|\nu_0\rangle=e^{-2r/a_X}\sqrt{8/\pi a_X^2}$, where 
$a_X$ is the 3D Bohr radius, this scattering reads, in 
$a_X^{-1}$ unit for momentum and 
$\xi_X=e^2a_X/L^2$ unit for scattering,
\begin{eqnarray}
\xi^\mathrm{dir}\left(^{-\v P'\ -\v P}_{\ \,\,\v P'\ \ \ \v 
P}\right)&\equiv& \xi^\mathrm{dir}_{\alpha_e}(Q)\nonumber\\
&=&\frac{2\pi}{Q}\left[g(\alpha_eQ)-g(\alpha_hQ)\right]^2\ ,
\end{eqnarray}
where $\alpha_e=1-\alpha_h=m_e/(m_e+m_h)$ while 
$g(q)=(1+q^2/16)^{-3/2}$. This shows that 
$\xi_{\alpha_e}^\mathrm{dir}(Q)$ reduces to zero for $Q=0$ and $Q$ 
infinite, while it stays equal to zero for $\alpha_e=1/2$, i.e., for 
equal electron and hole masses. Such a cancellation can be physically 
understood by noting that the exciton composite nature does not show 
up in a direct scattering, so that excitons basically behave as 
two classical dipoles, these dipoles being fully symmetrical when the 
electron and hole masses are equal.

\subsubsection{Exchange terms}

By contrast, the ``in'' exchange scattering cannot be calculated 
analytically. Its most compact expression appears to read  \cite{R.C.}
\begin{equation}
\xi^\mathrm{in}\left(^{-\v P'\ -\v P}_{\ \,\,\v P'\ \ \ \v 
P}\right)\equiv \xi^\mathrm{in}(\alpha_e\v Q-\alpha_h\v 
Q',-\alpha_e\v Q-\alpha_h\v Q')\ .
\end{equation}
The function $\xi^\mathrm{in}(\v u,\v v)$, which 
does not explicitly depend on the mass ratio, is precisely given by
\begin{eqnarray}
\xi^\mathrm{in}(\v u,\v v)=\sum_{\v q\neq\v 0,\v k,\eta=\pm 
1}V_{\v q}\hspace{3cm}\nonumber\\
\langle\nu_0|\v k+\frac{\v v+\v q}{2}\rangle\langle\nu_0|\v 
k-\frac{\v v+\v q}{2}\rangle\langle\v k+\frac{\v u+\eta\v 
q}{2}|\nu_0\rangle\nonumber\\
\times\left[\langle\v k-\frac{\v u+\eta\v 
q}{2}|\nu_0\rangle-\langle\v k-\frac{\v u-\eta\v 
q}{2}|\nu_0\rangle\right]\ ,
\end{eqnarray}
so that $\xi^\mathrm{in}(\v u,\v v)=\xi^\mathrm{in}(-\v u,\v v)=\xi^\mathrm{in}(-\v u,-\v v)$.

\subsubsection{Effective scattering}

Since a $(\v P'\leftrightarrow -\v P')$ exchange in Eq.(5) amounts to 
change $\v Q$ into $\v Q'$, the effective scattering for excitons 
going from states $(\nu_0,\v P)$,$(\nu_0,-\v P)$ to states $(\nu_0,\v 
P')$,$(\nu_0,-\v P')$, ends by reading as
\begin{eqnarray}
\xi^\mathrm{eff}\left(^{-\v P'\ -\v P}_{\ \,\,\v P'\ \ \ \v 
P}\right)\equiv\xi_{\alpha_e}^\mathrm{eff}(\v Q,\v 
Q')\hspace{3cm}\nonumber\\
=\left[\xi_{\alpha_e}^\mathrm{dir}(Q)-\xi^\mathrm{in}(\alpha_e\v 
Q-\alpha_h\v Q',-\alpha_e\v Q-\alpha_h\v Q')\right]\nonumber\\
+ [\v Q\leftrightarrow\v Q'],
\end{eqnarray}
where $\v Q$ and $\v Q'$ are the momentum transfers defined in Eq.(5) in terms of the 
exciton initial and final momenta $(\v P,\v P')$.

We have seen that, for equal electron and hole masses, i.e., for $\alpha_e=1/2$, the 
direct Coulomb scattering cancels. Due to the symmetry properties of
$\xi^\mathrm{in}(\v u,\v v)$,  the 
effective exciton-exciton scattering then reduces to one term only
\begin{eqnarray}
\xi_{1/2}^\mathrm{eff}(\v Q,\v Q')&=&-2\xi^\mathrm{in}\left((\v Q-\v 
Q')/2,(-\v Q-\v Q')/2\right)\nonumber\\
&=&-2\xi^\mathrm{in}(\v P',\v P)\ .
\end{eqnarray}
In the other limit, i.e., when the hole mass is infinite, $\alpha_e$ 
is equal to zero. Due to the symmetry properties of $\xi^\mathrm{in}(\v u,\v v)$, this effective scattering is then given by
\begin{eqnarray}
\xi_0^\mathrm{eff}(\v Q,\v 
Q')&=&\left[\xi_0^\mathrm{dir}(Q)-\xi^\mathrm{in}(-\v Q',-\v 
Q')\right]\ +\ [\v Q\leftrightarrow\v Q']\nonumber\\
&=&\left[\xi_0^\mathrm{dir}(Q)-\xi^\mathrm{in}(\v Q,\v 
Q)\right]\ +\ [\v Q\leftrightarrow\v Q']\
\end{eqnarray}

\section{Possible cancellation of the effective exciton-exciton scattering}

\subsection{Previous work}

In a previous work \cite{R.C.}, we already calculated the ``in'' 
exchange Coulomb scattering in the particular case of initial 
excitons having equal momenta in the laboratory frame. This 
configuration corresponds to $\v P=\v 0$, i.e., $\v Q'=-\v Q$. 
According to Eq.(24), this means that we already calculated
\begin{equation}
\xi^\mathrm{in}(\v Q,[\alpha_h-\alpha_e]\v Q)=\xi_{\alpha_e}^\mathrm{in}(Q)\ .
\end{equation}
By contrast with the direct scattering which stays positive, this 
``in'' exchange scattering is negative for small $Q$ but turns 
positive when $Q$ gets large. More precisely, for zero momentum 
transfer, $\v Q=0$, i.e., when the two ground state excitons $(\nu_0,\v 
0)$ stay in the same $(\nu_0,\v 0)$ state, 
$\xi_{\alpha_e}^\mathrm{in}(0)$ in 2D is equal to \cite{Dupertuis} 
$-(4\pi-315\pi^3/1024)\simeq -3.0$ whatever the carrier masses are, 
while $\xi_{\alpha_e}^\mathrm{in}(Q)$ turns positive for a momentum 
transfer $Q_{\alpha_e}^{(0)}$ which slightly varies with $\alpha_e$,
\begin{equation}
\xi_{\alpha_e}^\mathrm{in}(Q)=0,\ \ \mathrm{for}\ \ Q=Q_{\alpha_e}^{(0)}\ .
\end{equation}
For infinite hole mass, this momentum transfer is equal to 
$Q_0^{(0)}\simeq 2.7$.

In this previous work, we also calculated 
$\xi_{\alpha_e}^\mathrm{eff}(\v Q,\v Q')$ for $\v P=\v 0$. Since  $\v P=\v 0$
corresponds to $\v Q'=-\v Q$, we thus also know
\begin{equation}
\xi_{\alpha_e}^\mathrm{eff}(\v Q,-\v Q)\equiv \xi_{\alpha_e}^\mathrm{eff}(Q)\ .
\end{equation}
This effective scattering was found to cancel for a momentum 
transfer which slightly depends on mass ratio,
\begin{equation}
\xi_{\alpha_e}^\mathrm{eff}(Q)=0\ \ \mathrm{for}\ \ Q=Q_{\alpha_e}^\ast\ .
\end{equation}
  $Q_{\alpha_e}^\ast$ varies from $Q_{1/2}^\ast\simeq3.1$ to 
$Q_0^\ast\simeq3.9$ when the hole mass increases, the effective 
scattering $\xi_{\alpha_e}^\mathrm{eff}(Q)$ staying very close to zero for $Q$ larger than 
$Q_{\alpha_e}^\ast$. However, since energy conserving scattering when 
$\v P=\v 0$ imposes $\v P'=\v 0$, i.e., $Q=0$, such a momentum 
transfer $Q_{\alpha_e}^\ast$ for cancellation of $\xi^\mathrm{eff}$ 
is of no physical relevance because it corresponds to process in 
which energy is not conserved.

  When $P=0$, the scattered state having the same energy as the initial state corresponds 
to $P'=P$: it thus reduces to the initial state, which makes the 
$P=0$ initial configuration not so much of interest. In order to 
consider physically relevant energy conserving configurations, we must
extend our previous calculations to finite initial momentum 
difference.

Since the direct Coulomb scattering is analytically known, this means 
that we have to numerically calculate the ``in'' exchange Coulomb 
scattering for arbitrary initial momenta. From it, we will then determine, for various scattered angles $\theta$ and mass 
ratios $\alpha_e$, the value of the initial momentum difference 
for which the effective scattering defined in Eq.(26) cancels when 
$P=P'$, i.e., when energy is conserved. Let us call 
$P^{(0)}_{\alpha_e}(\theta)$ the $P$ value for which such cancellation occurs,
\begin{equation}
\xi^\mathrm{eff}_{\alpha_e}(\v Q,\v Q')=0\ \ \mathrm{for}\ \ 
P=P'=P^{(0)}_{\alpha_e}(\theta)\ .
\end{equation}
The resulting $\theta$ dependence of this half initial momentum difference is given in Figs 9 and 11 for infinite hole mass and equal electron and hole masses. Let us now derive these results more in details.

\subsection{Infinite hole mass}

Actually, our previous work \cite{R.C.} in which we only considered 
$P=0$, is enough to get the effective scattering for a general $(\v 
P,\v P')$ configuration when the hole mass is infinite, i.e., when 
$\alpha_e=0$. Indeed, the ``in'' exchange scattering in Eq.(24) then 
reduces to $\xi^\mathrm{in}(-\v Q',-\v Q')$ which is nothing but 
$\xi_0^\mathrm{in}(Q')$, according to Eq.(29). Consequently, the 
effective scattering, given in Eq.(28) reads as
\begin{eqnarray}
\xi^\mathrm{eff}_0(\v Q,\v Q')
&=&[\xi_0^\mathrm{dir}(Q)-\xi_0^\mathrm{in}(Q)]+[Q\leftrightarrow 
Q']\nonumber\\
&=&\frac{1}{2}[\xi_0^\mathrm{eff}(Q)+\xi_0^\mathrm{eff}(Q')]\ .
\end{eqnarray}
Due to Eq.(32), this effective scattering obviously cancels for 
$Q=Q'=Q_0^\ast$. It also cancels for configurations having different momentum transfers, $Q$ and $Q'$ then being on both sides of $Q_0^\ast$. If we now
restrict to processes in which energy is conserved, $Q^2+Q'^2=4P^2$, 
this cancellation occurs when half the initial momentum difference $P$ is
larger than a threshold value $P_0^{\ast}$ which precisely 
corresponds to equal momentum transfers $Q=Q'=Q_0^{\ast}$, so that 
this initial momentum threshold corresponds to
\begin{equation}
P_0^{\ast}=Q_0^{\ast }/\sqrt{2}\simeq 2.77.
\end{equation}
 \begin{figure}[t]
\includegraphics[scale=0.36]{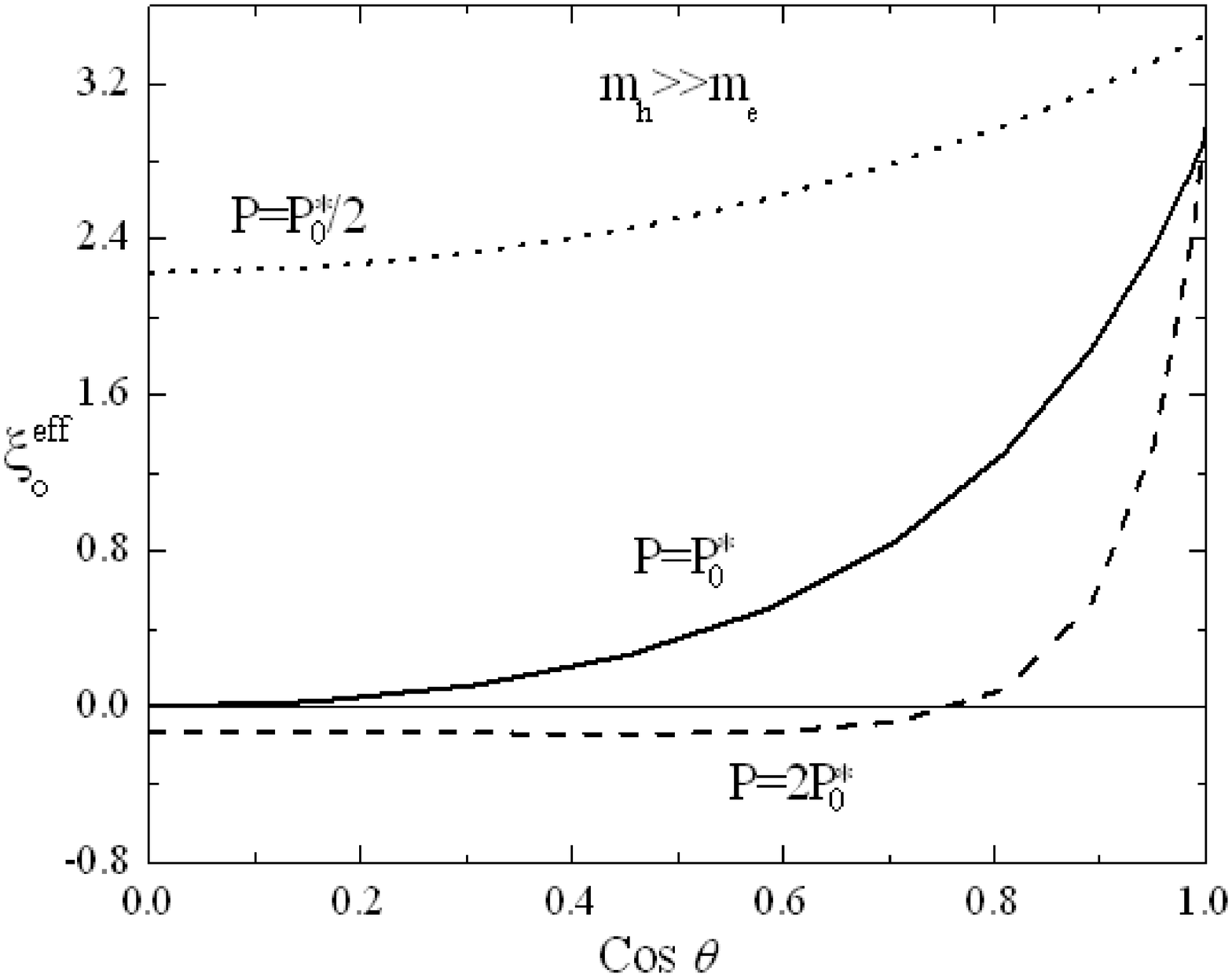}
\caption{Effective scatterings  $\xi^\mathrm{eff}_{\alpha_e=0}$ in 
the case of infinite hole mass, $m_e/m_h=0$, for three different 
values of  half the initial momentum difference $P$, namely, 
$P_0^{\ast}/2, P_0^{\ast}, 2P_0^{\ast}$, where $P_0^{\ast}$ is the 
threshold value of P above which the effective exciton-exciton 
scattering can cancel.}
\end{figure}

Since $\v P.\v P'=0$ for $Q=Q'$, cancellation at threshold occurs for 
scattering in the perpendicular direction ($\cos\theta=0$). For $P$ 
larger than this threshold value $P_0^\ast$, the angle $\theta$ 
between $\v P$ and $\v P'$ when cancellation occurs, decreases, the 
effective scattering staying however small for all configurations. 
This means that, for $P\gg P_0^{\ast}$, sizeable scatterings exist in 
the forward direction only ($\cos\theta\simeq\pm 1$), i.e.,  
$\theta\simeq (0\ \mathrm{or}\ \pi)$. This behavior is shown in Fig.7, which gives 
the effective scatterings for three different values of $P$ taken 
below, at and above threshold, namely, $P=P_0^{\ast}/2$, $P_0^{\ast}$ 
and $2P_0^{\ast}$. We in particular see that, for $P=2P_0^\ast$, 
cancellation occurs for $\cos\theta\simeq 0.75$.
 \begin{figure}[t]
\includegraphics[scale=0.37]{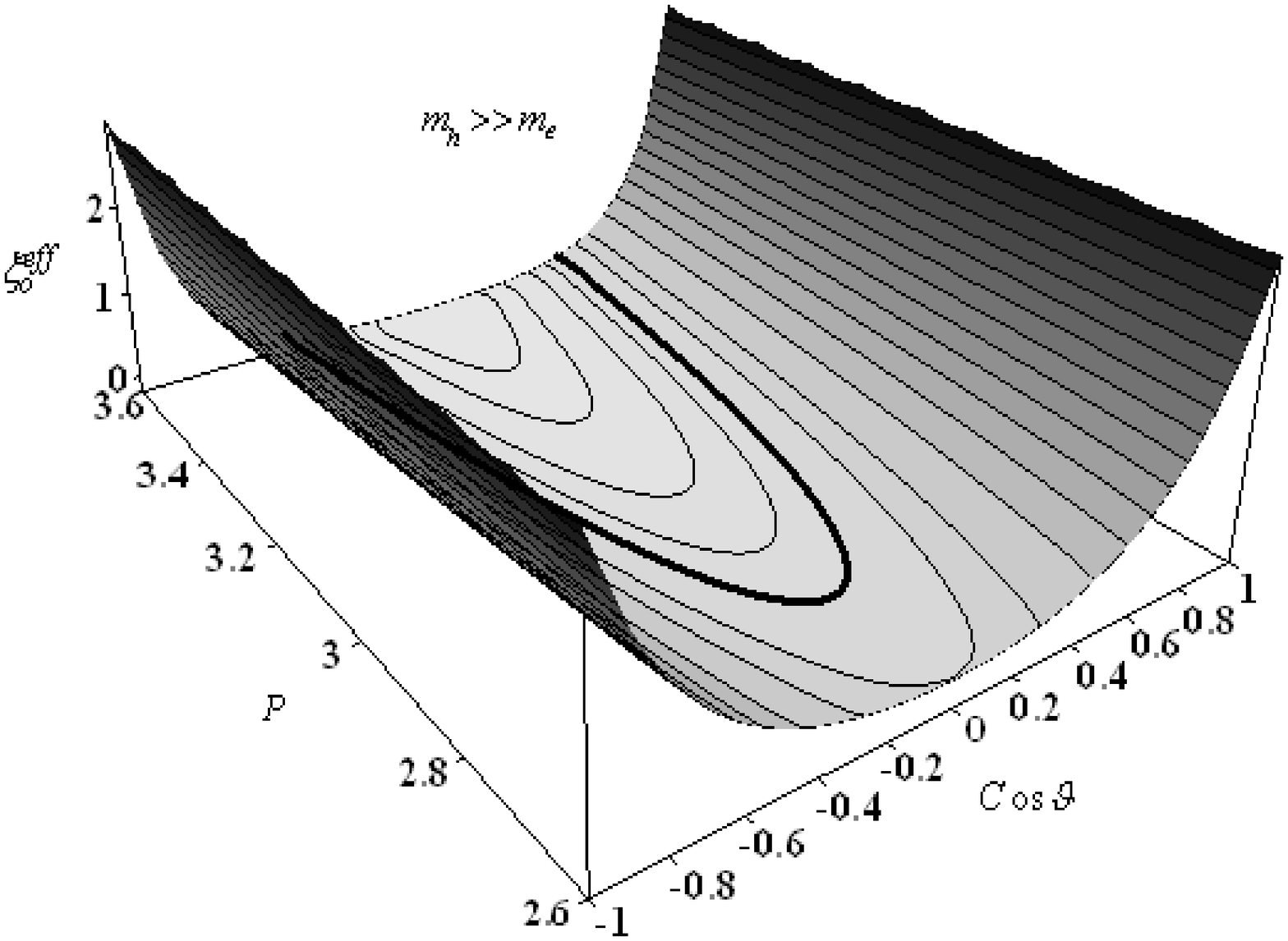}
\caption{Effective scatterings  $\xi^\mathrm{eff}_{\alpha_e=0}$ for 
energy conserving processes as a function of half the initial 
momentum difference P and the angle $\theta$ between initial and 
scattered momenta when the hole mass is infinite. The full lines 
correspond to constant $\xi^\mathrm{eff}_{\alpha_e=0}$, the heavy one 
corresponding to $\xi^\mathrm{eff}_{\alpha_e=0}=0$.}
\end{figure}

 \begin{figure}[t]
\includegraphics[scale=0.35]{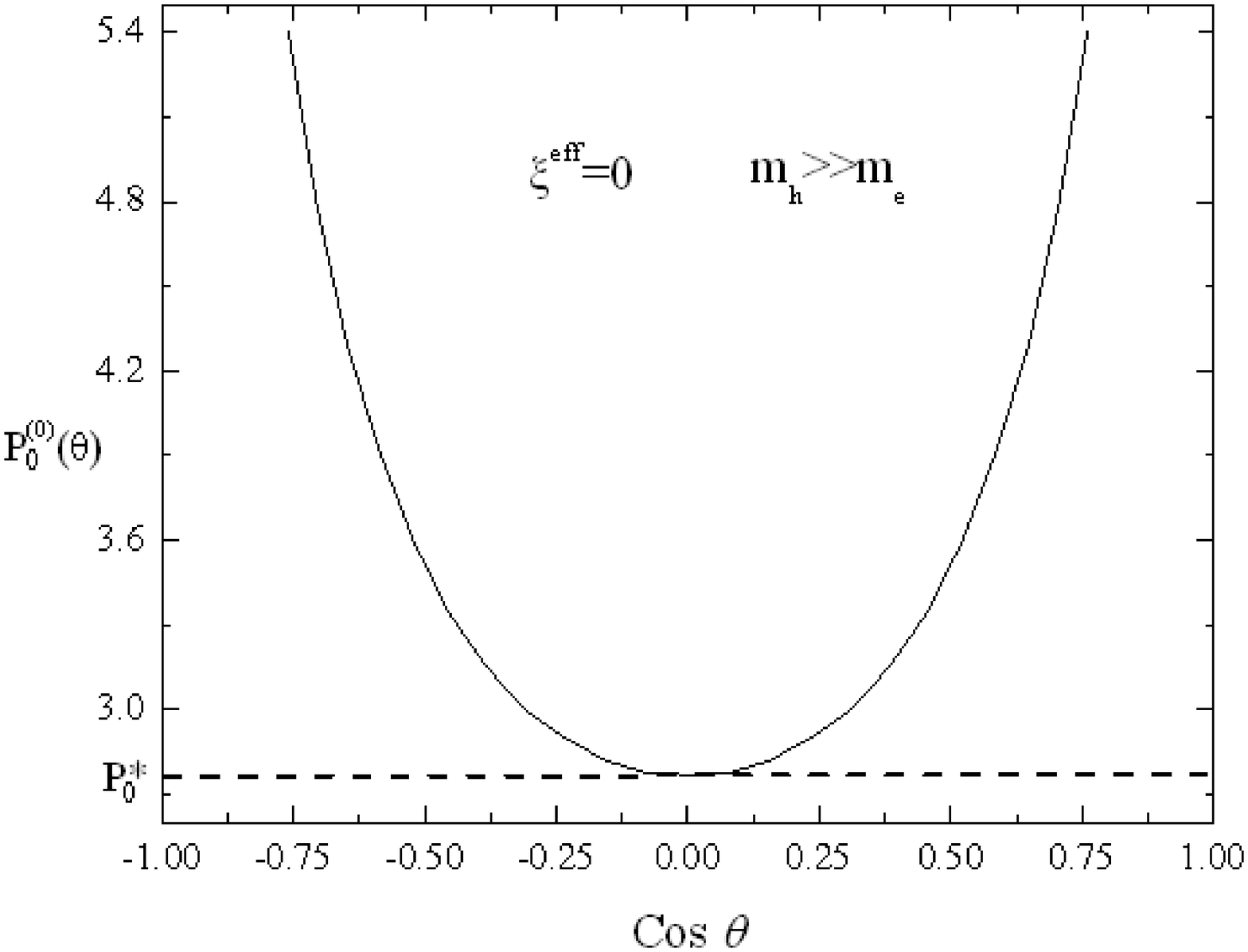}
\caption{Initial half exciton momentum difference 
$P^{(0)}_{\alpha_e=0}(\theta)$ for effective scattering cancellation 
as a function of the angle between initial and scattered momenta when the 
hole mass is infinite. The dashed line corresponds to the treshold 
value $P_0^{\ast}\simeq 2.77$ above which cancellation can occur [see Eq.(35)].}
\end{figure}

Using Eq.(34), it is possible to calculate the effective scattering 
as a function of half the momentum difference $P$ and the angle 
$\theta$ between scattered momenta.
In Fig.8, the effective scattering landscape for energy conserving 
configurations is shown for a large range of $P$ values. The heavy 
line corresponds to zero effective scattering.
Sections with planes where $P$ is constant give curves similar to the 
ones of Fig.7. Since the direct Coulomb scattering $\xi^\mathrm{dir}$ is always positive [see Eq.(23)], Fig.8 shows how compensation 
between direct and exchange Coulomb processes evolves to produce 
cancellation. The  $\xi_0^\mathrm{eff}=0$ heavy line 
separates the effective scattering surface into two regions; when 
$\xi_0^\mathrm{eff}<0$, the exchange term dominates over the direct 
one as in the small momentum limit.

In order to better characterize this cancellation effect, we also 
show in Fig.9 the curve $P^{(0)}_{0}(\theta)$ [defined in Eq.(33)] 
where $\xi_0^\mathrm{eff}$ cancels. Among all possible final states 
satisfying energy and momentum conservations, this curve selects the 
forbidden ones at first order in Coulomb process.
In agreement with Fig.7, the minimum value of $P^{(0)}_0(\theta)$ 
occurs at threshold $P^{(0)}_0(\pi/2)=P_0^\ast$ for $\cos\theta=0$, 
while for $2P_0^{\ast}$ it occurs for 
$\cos\theta\simeq \pm0.75$.

\subsection{Equal electron and hole masses}

We now turn to the other limit, i.e., equal electron and hole masses. 
As seen from Eq.(27), the effective scattering in the forward 
direction, $\v P=\v P'$, reduces to $-2\xi^\mathrm{in}(\v P,\v P)$ 
which is nothing but $-2\xi_0^\mathrm{in}(P)$, due to Eq.(29). This 
effective scattering is thus found to cancel for $P$ equal to 
$Q^{(0)}_0\simeq 2.7$ [see Eq.(30)], so that the momentum 
$P^{(0)}_{1/2}(\theta=0)$ for cancellation of the effective 
scattering, defined in Eq.(33), is equal to $Q_0^{(0)}$.

Through a numerical calculation of the scattering $\xi^\mathrm{in}(\v P',\v P)$ 
given in Eq.(25), when energy is conserved, i.e., for $P=P'$, we can 
get the effective scattering $\xi_{1/2}^\mathrm{eff}$ as a function 
of half the momentum difference $P$ and scattering angle $\theta$ and 
determine where it cancels when energy is conserved. By comparing Figs.(8) and (10), we see that the behaviours of $\xi^\mathrm{eff}_{\alpha_e=0}$ and $\xi^\mathrm{eff}_{\alpha_e=1/2}$ for 
energy conserving configurations are quite different. Most strikingly, the curve $P^{(0)}_{1/2}(\theta)$ for 
cancellation, shown more in details in Fig.11, has a curvature opposite to the one 
for infinite hole mass, shown in Fig.9:
scattering in the perpendicular direction, $\v P'.\v P=0$, is found 
to cancel for a $P$ value $P^{(0)}_{1/2}(\theta=\pi/2)\simeq3.66$ 
which is larger than the value in the forward direction 
$P^{(0)}_{1/2}(\theta=0)\simeq 2.7$. By contrast, the minimum value of $P^{(0)}_{\alpha_e=0}(\theta)$, when the hole mass is infinite, is reached for $\theta=0$.

Such different behaviors of $P^{(0)}_0(\theta)$ and 
$P^{(0)}_{1/2}(\theta)$ can look quite strange at first, because, from the behaviors of
these two extreme mass ratios, $m_e/m_h=0$ and 
$m_e=m_h$, we expect, by continuity, to go through a value of 
$\alpha_e=m_e/(m_e+m_h)$ for which $P^{(0)}_{\alpha_e}(\theta)$ would 
stay constant and equal to zero when changing the angle $\theta$ 
between the initial and final state exciton momenta: for half initial 
momentum difference equal to this constant value, the effective 
scattering would then stay equal to zero in all directions.
  \begin{figure}[t]
  \hspace{-1cm}
\includegraphics[scale=0.36]{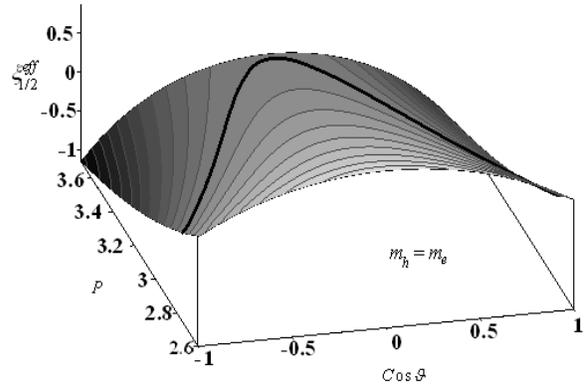}
\caption{Effective scatterings  $\xi^\mathrm{eff}_{\alpha_e=1/2}$ for 
energy conserving processes as a function of half the initial 
momentum difference P and the angle $\theta$ between initial and 
scattered momenta when the electron and hole masses are equal. The 
full lines correspond to constant $\xi^\mathrm{eff}_{\alpha_e=1/2}$, 
the heavy one corresponding to $\xi^\mathrm{eff}_{\alpha_e=1/2}=0$.}
\end{figure}
 \begin{figure}
\includegraphics[scale=0.35]{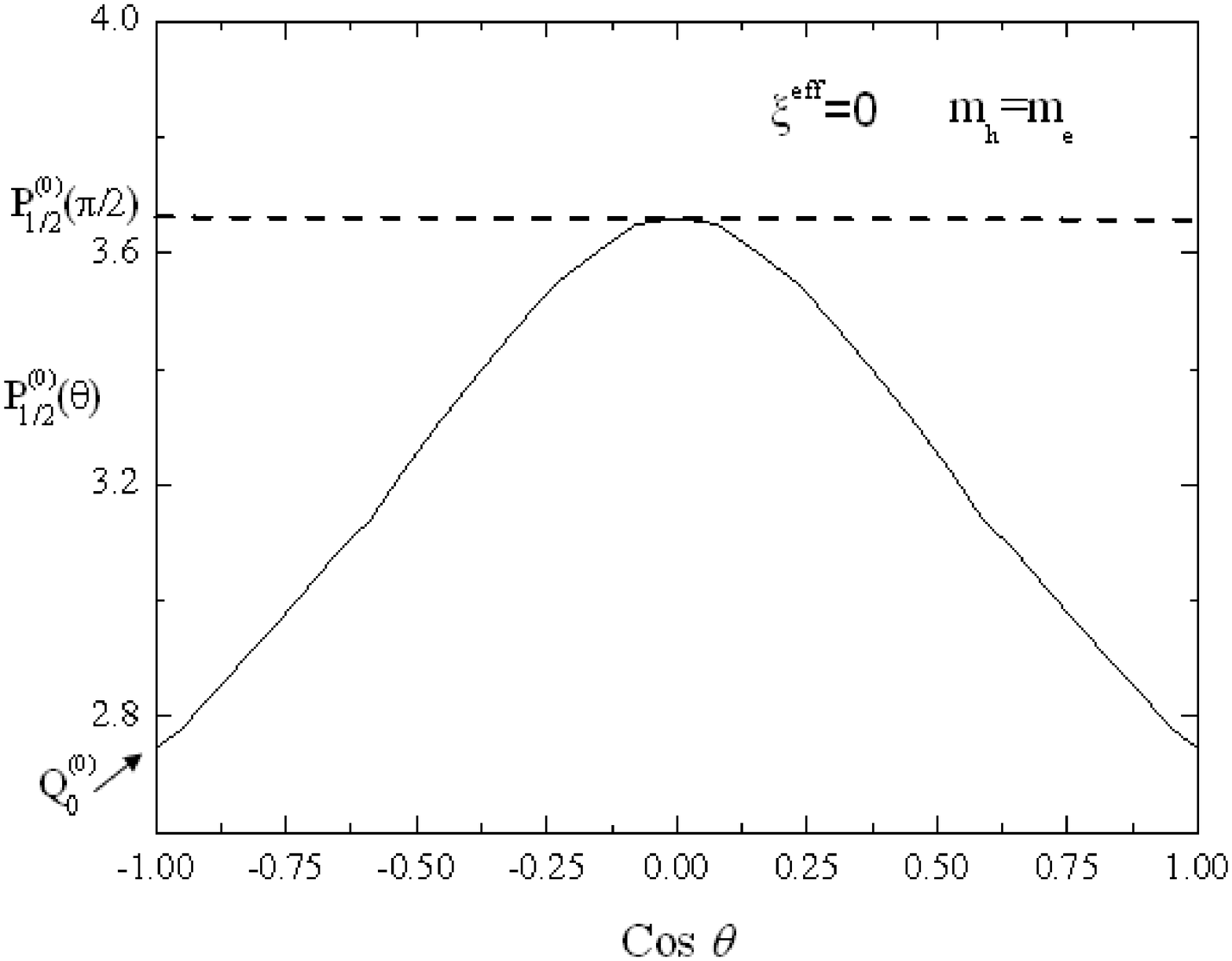}
\caption{Initial half exciton momentum difference 
$P^{(0)}_{\alpha_e=1/2}(\theta)$ for effective scattering 
cancellation as a function of the angle between initial and scattered 
momenta for equal electron and hole masses.}
\end{figure}

Let us now determine the value of
the electron-to-hole mass ratio which separates these two different 
regimes of curvature, in order to characterize more in details the 
magic initial configuration having a zero effective scattering in all directions.

\subsection{Arbitrary mass ratio}

We start from the general form of $\xi^\mathrm{eff}\left(^{-\v P'\ 
-\v P}_{\ \,\,\v P'\ \ \ \v P}\right)$ given in Eq.(26) which, 
together with Eqs.(23) and (25),  gives the effective scattering as a 
function of the mass ratio $\alpha_e$ and the momentum transfers $(\v 
Q,\v Q')$, these momentum transfers being related to the exciton 
initial and final momenta $(\v P,\v P')$ through Eq.(5). We again 
consider energy-conserving processes, i.e., processes such that 
$P=P'$. The effective scattering then depends on the initial half 
momentum difference $P$, the angle $\theta$ between $\v P$ and $\v 
P'$ and the mass ratio $\alpha_e$. Let us call it 
$\xi^\mathrm{eff}_{\alpha_e}(P,\theta)$.

In order to better see if an initial state with a zero effective 
scattering in all scattered directions can exist, we first look for 
values of the initial half momentum difference $\overline{P}$, and 
mass ratio $\overline{\alpha}_e$, for which the effective scattering 
cancels in the forward and perpendicular directions. This corresponds 
to look for $\overline{P}$ and $\overline{\alpha}_e$ such that
\begin{equation}
\xi^\mathrm{eff}_{\overline{\alpha}_e}(\overline{P},0)=\xi^\mathrm{eff}_{\overline{\alpha}_e}(\overline{P},\pi/2)=0\ .
\end{equation}
 
 We find that this happens for $\overline{P}\simeq 3.3$ and $\overline{\alpha}_e\simeq 
0.32$, which corresponds to a hole mass value $m_h$ of the order of 
$2.08m_e$.
Figure 12 shows the half momentum difference 
$P^{(0)}_{\overline{\alpha}_e}(\theta)$ for effective scattering 
cancellation when the mass ratio is equal to $\overline{\alpha}_e$. We see that $P^{(0)}_{\overline{\alpha}_e}(\theta)$ is indeed equal to $\overline{P}$ for $\theta=0$ and 
$\theta=\pi/2$ but does not stay exactly equal to $\overline{P}$ when changing 
$\theta$: this half momentum difference actually shows a very small 
oscillation, crossing the $P=3.33$ value in four 
points when $\cos\theta$ varies from $-1$ to $+1$. Although this 
oscillation is very small, it actually rules out a far more 
striking behavior, with an effective scattering staying \emph{exactly} 
equal to zero for all scattering angles $\theta$.

\begin{figure}[t]
\includegraphics[scale=0.35]{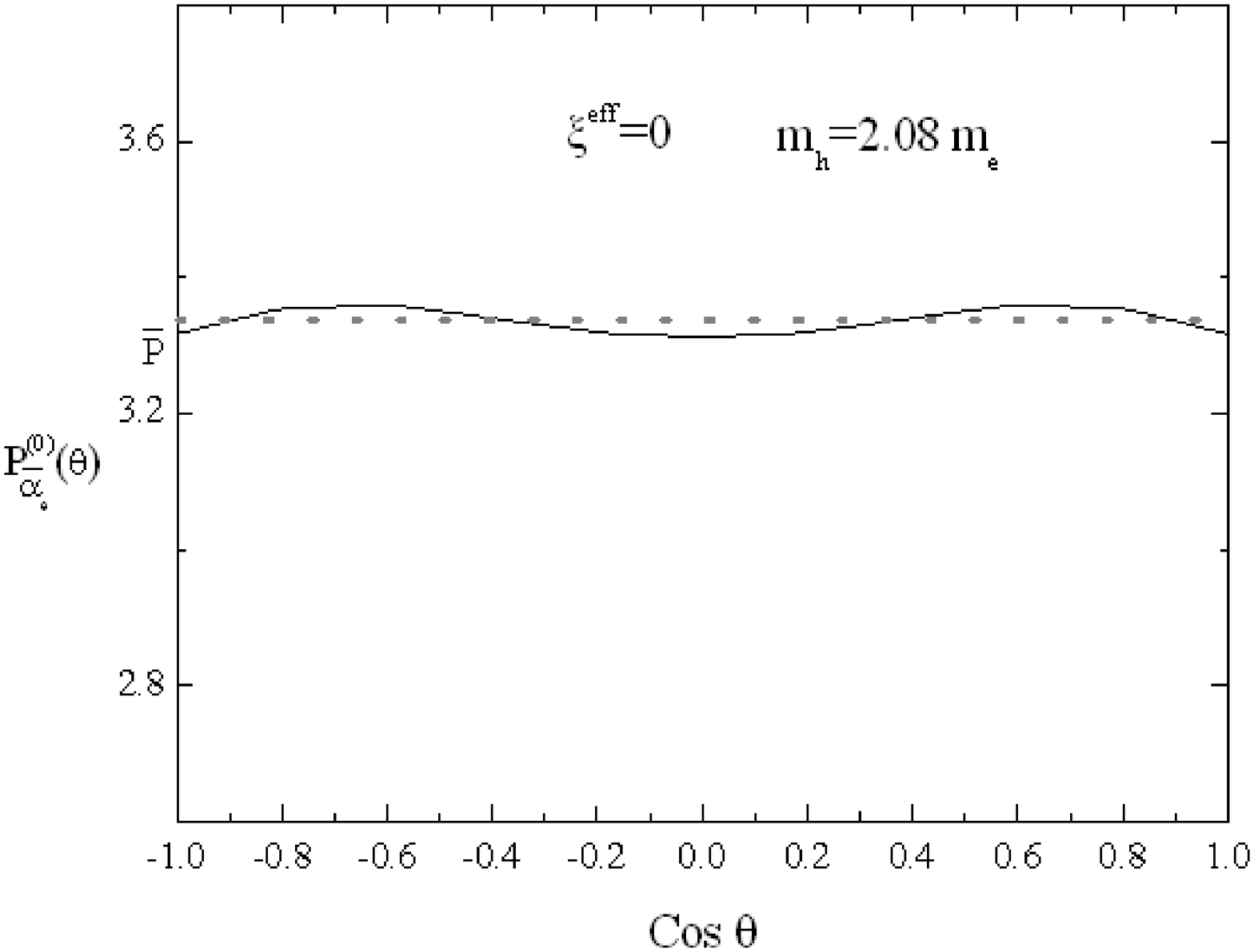}
\caption{Initial half exciton momentum difference 
$P^{(0)}_{\alpha_e}(\theta)$ for effective scattering 
cancellation as a function of the angle between initial and scattered 
momenta when $m_h=2.08m_e$, i.e., when $\alpha_e=\overline{\alpha}_e=0.32$. The dotted line, $P=3.33$, is to guide the eyes for the weak oscillation of $P^{(0)}_{\overline{\alpha}_e}(\theta)$, the value of $P^{(0)}_{\overline{\alpha}_e}(\theta)$ being exactly equal to $\overline{P}\simeq 3.3$ for $\cos\theta=(0,\pm 1)$}
\end{figure}

In order to better characterize this fundamental collapse of the 
effective exciton-exciton scattering, we have performed calculations 
for values of $\alpha_e$ very close to $\overline{\alpha}_e$, namely, 
$m_h=(2.08\mp0.08)m_e$: the results are shown in Fig.13  and Fig.14 
respectively. In both cases, we find that there are very narrow 
ranges of $P$ values for which scattering cancellation occurs 
$(\Delta P/P\simeq 0.03)$.

 \begin{figure}[t]
\includegraphics[scale=0.35]{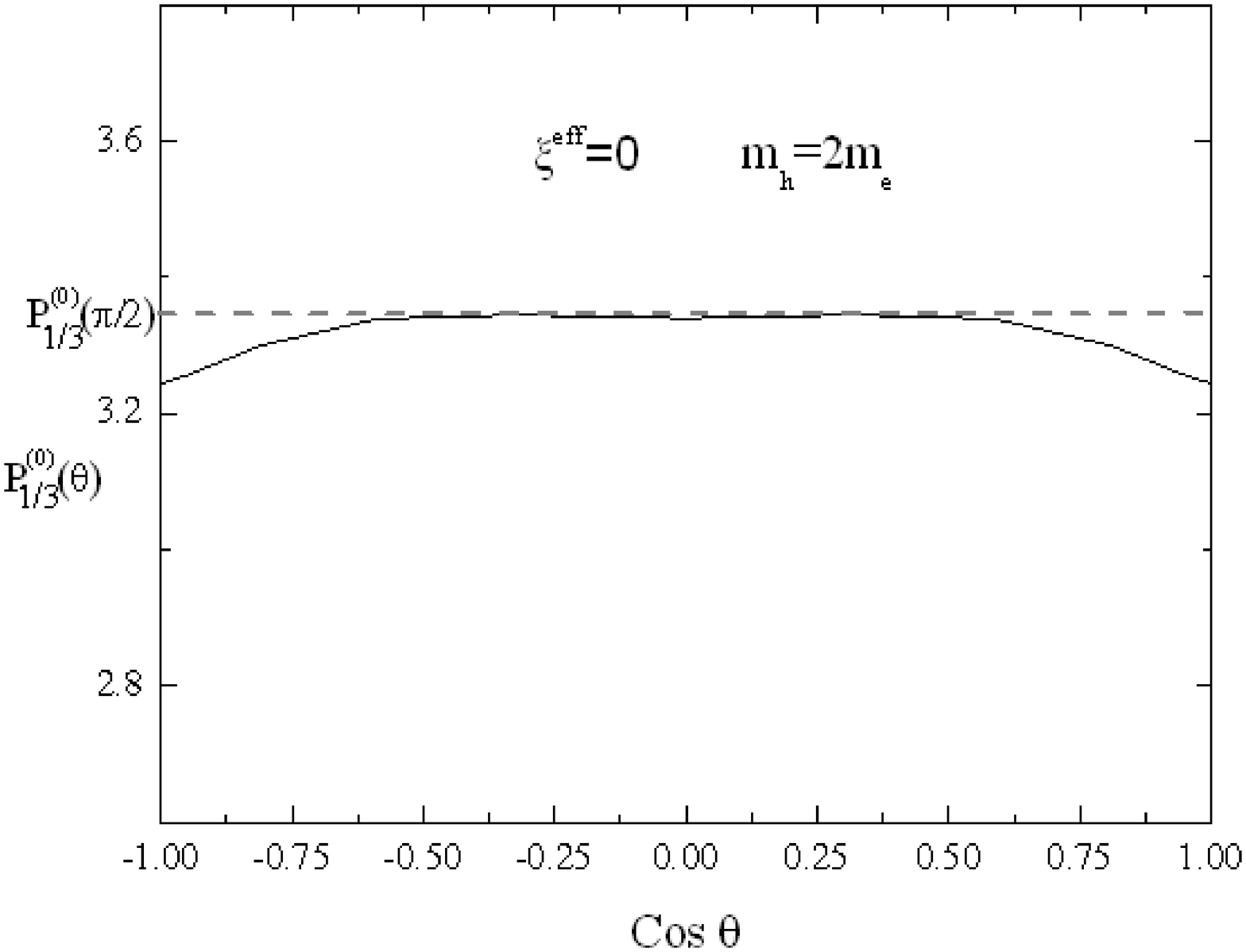}
\caption{Half initial exciton momentum difference 
$P^{(0)}_{\alpha_e=1/3}(\theta)$ for cancellation of the effective scattering, 
as a function of the angle between initial and scattered 
momenta when $m_h=2m_e$.}
\end{figure}

 \begin{figure}[t]
\includegraphics[scale=0.35]{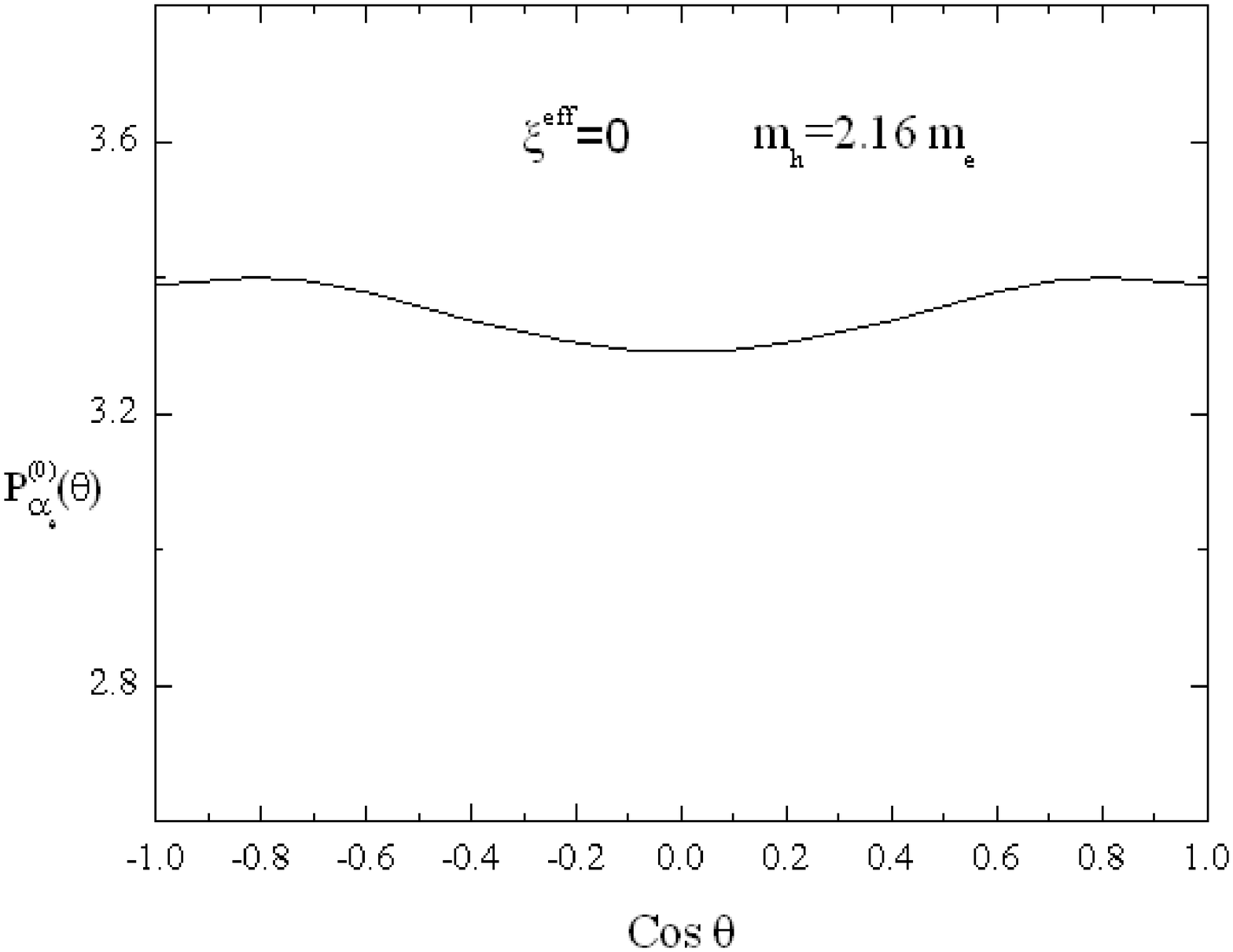}
\caption{Half initial exciton momentum difference 
$P^{(0)}_{\alpha_e}(\theta)$ for cancellation of the effective scattering, as 
a function of the angle between initial and scattered momenta when 
$m_h=2.16m_e$.}
\end{figure}

Of particular interest is the case $m_h=2m_e$ for which, when the 
initial half momentum difference $P$ is equal to 
$P^{(0)}_{1/3}(\pi/2)$ (see Fig.13), the effective scattering stays essentially equal 
to zero for all scattered directions $\theta$ between 
$\pi/4$ and $3\pi/4$. This rather large 
range of $\theta$ values in which energy conserving scattering is forbidden, 
could possibly help to evidence this highly non intuitive fundamental collapse. 
We see that, for this $P$ value, non-zero scattering exists for 
$\cos \theta$ very close to $\pm1$, i.e., in the forward 
direction only. 

The $m_h/m_e\simeq2.08$ mass ratio, which fundamentally separates two different regimes of curvature for $P^{(0)}_{{\alpha}_e}(\theta)$, can be physically seen as the precise
value which separates positronium-like 
excitons with $m_e\simeq m_h$ (Fig.8) from hydrogen-like 
excitons with $m_e\ll m_h$ (Fig.10).

It is clear that the cancellation of the exciton-exciton effective scattering requires exciton initial momenta far larger than photon momenta, i.e., the momenta of photocreated excitons. As a result, this quite remarkable collapse seems hard to experimentally evidence in a direct way through excitons created by resonant photons. Nevertheless, its existence could have consequences in physical properties related to exciton-exciton scatterings for systems having excitons with high kinetic energy as possibly produced by non-resonant photons.

For example to explain the rise time observed in time resolved luminescence of GaAs/AlGaAs/GaAs(001) 2D multi quantum well, the authors of Ref.[39] speculate that exciton-exciton scattering plays a dominant role with respect to dephasing effects linked to either disorder or acoustic phonons. It is legitimate to then expect  strongly different behavior when such dominant mechanism is hampered by the present cancellation effect.

We wish to stress that, although rather large, the half initial exciton momentum difference $P^{(0)}_{\alpha_e}$ for which scattering cancels, stays below the exciton ionization threshold, for all mass ratios considered here, as can be seen from Fig.4. Consequently, the asumption of excitons staying in their fundamental ground state, under which these calculations are made, is fully valid.

The present work considers quasi two-dimensional quantum wells ($L\ll a_X$) with hole-to-electron mass ratio $m_h /m_e$ close to 2 (Fig.13). This is easily fulfilled in high quality GaAs/AlAs/GaAs(001)  with electron mass $m_e $=0.067 and hole mass in the parallel direction $m_{hh}^{//} $=0.110. The 3D Bohr radius being $a_{X} $=17.5nm, the 2D Rydberg energy $4R_{yd}$ is of the order of 12meV while the ionization threshold for center-of mass energy is of the order of 7.8meV (see Fig.4). Moreover, being the difference between the lowest (n=0) and the first excited state (n=1), for 2D excitons, $\Delta E\approx 3.55R_{yd}$ we can check that the exciton kinetic energy is lower than the transition energy $\Delta E$.
Moreover, high quality samples are necessary to minimize dephasing induced by interface disorder. A very accurate control of the sample temperature is also needed to get rid of exciton-acoustic phonon interaction. A precise control of the heavy-light hole splitting energy, induced by small difference in the lattice parameters between GaAs and AlAs materials (compressive strain) is also required. Note that this splitting must be added to the quantum confinement energy (for GaAs the z-masses are $m_{hh}^{z}=0.530, m_{lh}^{z}=0.08$ ).

All this tends to show that, altought the experimental observation of the exciton-exciton scattering cancellation is going to require a rather sophisticated tailoring of the sample as well as non-linear optical experiments designed in an unusual way, the observation of such an unexpected effect which is directly linked to the existence of fermionic components in excitons, does not seem out of reach and seems to us a challenge of physical interest.

\section{Conclusion}

Through the composite exciton many-body framework recently proposed by 
Combescot and coworkers, we here study the effective scattering ruling 
the time evolution of two excitons 
at first order in Coulomb interaction (Born approximation) when these two excitons are in the same relative motion state but have different initial momenta. We mainly look for energy conserving configurations in which this effective scattering cancels. 

(i) We show that the possible cancellation of this effective scattering is fundamentally due to the exciton composite nature:  such a cancellation always occurs for elementary fermions but never occurs for elementary bosons. In the case of composite-boson excitons made of two fermions, cancellation can occur but above a momentum threshold only.

(ii) The effective scattering ruling the time evolution of two excitons
shows a strong dependence on the electron-to-hole mass ratio. For $m_h/m_e$ close to $2$, which can be seen as a boundary between 
hydrogen-like excitons for which $m_e\ll m_h$ and positronium-like 
excitons for which $m_e\simeq m_h$, we find a quite remarkable cancellation of 
this effective scattering over a large range of 
scattered directions, $\theta\simeq(\pi/4,3\pi/4)$, but a very narrow range of initial exciton momentum difference, 
$\Delta P/P\simeq 0.03$.

\section*{Acknowledgments}
We wish to thank M.A. Dupertuis for a valuable help in the numerical 
part of this work. One of us (M.C.) wishes to thank the Italian CNR 
"Istituto dei Sistemi Complessi" for an invitation, through the Short 
Term Mobility program, during which most of this work was completed.

\end{document}